\documentstyle[11pt,aaspp4]{article}
\def\gax{\mathrel{\raise.3ex\hbox{$>$}\mkern-14mu\lower0.6ex\hbox{$\sim$}}}
\def\lax{\mathrel{\raise.3ex\hbox{$<$}\mkern-14mu\lower0.6ex\hbox{$\sim$}}}
\def\gtorder{\mathrel{\raise.3ex\hbox{$>$}\mkern-14mu
             \lower0.6ex\hbox{$\sim$}}}
\def\ltorder{\mathrel{\raise.3ex\hbox{$<$}\mkern-14mu
             \lower0.6ex\hbox{$\sim$}}}

\input psfig

\begin{document}

\title{Gravitational Lens Time Delays in CDM}

\author{C.S. Kochanek}
\affil{Harvard-Smithsonian Center for Astrophysics, 
       60 Garden Street, Cambridge, MA 02138}
\affil{email: ckochanek@cfa.harvard.edu}

\def\kbar{\langle \kappa \vphantom{R_1^2}\rangle}
\def\kbarp{\langle \kappa \vphantom{R_1^2}\rangle}
\def\ra{R_1}
\def\rb{R_2}

\begin{abstract}
In standard CDM halo models, the time delay of a gravitational lens
is determined by the cold baryon mass fraction, $f_b=\Omega_{b,cold}/\Omega_0$,
of the visible galaxy relative to the overall halo.  The observed time delays in 
PG1115+080, SBS1520+530, B1600+434 and HE2149--2745 give Hubble constants
consistent with the HST Key Project value of $H_0=72 \pm 8$~km/s~Mpc
only if $f_b \gtorder 0.2$ (1-sided 68\% confidence), which is larger than
the upper bound of $f_{b,max}=\Omega_b/\Omega_0=0.15\pm0.05$ estimated from the
CMB.  If all available baryons cool and $f_b=f_{b,max}$ then the time delays
imply $H_0=65 \pm 6$~km/s~Mpc (95\% confidence).
If local inventories of cold baryons, $f_b\simeq 0.013/h_{70}$,
are correct, then $H_0=52\pm 6$~km/s~Mpc and the halo parameters closely
match isothermal mass models.  Isothermal models are also consistent
with strong and weak lens studies, stellar dynamics and X-ray observations
on these scales, while significantly more centrally concentrated models
are not. There is a 
a conflict between gravitational lens time delays, the local distance
scale and standard CDM halo models.
\end{abstract}

\keywords{cosmology: gravitational lensing; cosmology: Hubble constant; dark matter}

\section{Introduction}

Kochanek~(\cite{Kochanek02a}) found that it was difficult to reconcile
the time delays measured for 5 simple, well-observed gravitational lenses with the local
distance scale given our expectation that galaxies have massive, extended
dark matter halos.  If the lens galaxies had constant
mass-to-light ($M/L$) ratios we found $H_0=71\pm6$~km/s~Mpc, which is 
consistent with the local estimate of $H_0=72\pm8$~km/s~Mpc by the HST Key 
Project (Freedman et al.~\cite{Freedman01}).  However, if the lenses had isothermal 
mass distributions (flat rotation curves), we found $H_0=48_{-4}^{+7}$~km/s~Mpc,
which is grossly inconsistent with the HST Key Project.  While the time delay 
lenses cannot distinguish between these two limiting mass distributions,
models of other lenses (e.g. Munoz, Kochanek \& Keeton~\cite{Munoz01}), stellar 
dynamical measurements (e.g. Rix et al.~\cite{Rix97}, Romanowsky \& Kochanek~\cite{Romanowsky99},
Gerhard et al.~\cite{Gerhard01}, Treu \& Koopmans~\cite{Treu02}), weak lensing 
(e.g. Guzik \& Seljak~\cite{Guzik02}) and X-ray (e.g. Fabbiano~\cite{Fabbiano89},
Lowenstein \& White~\cite{Lowenstein99}) measurements all suggest that the 
isothermal mass distributions are correct.  In this study we will show that
standard cold dark matter (CDM) halo models closely resemble the isothermal 
models on these scales, which implies there is a conflict between 
the local distance scale, gravitational lens time delays and CDM halo models.

While the Kochanek~(\cite{Kochanek02a}) results provided evidence for a real
conflict given the considerable observational evidence that lens galaxies must
have extended, massive dark matter halos, the link to a problem with CDM halo
models was qualitative because the study lacked a quantitative, theoretical
prediction for the time delays expected from CDM halos.  One barrier to 
making such predictions was that we lacked a clear understanding of which
features of lens mass distributions control time delays.  While global 
degeneracies due to the addition of constant mass density sheets 
(e.g. Falco, Gorenstein \& Shapiro~\cite{Falco85}, Gorenstein, Falco \& 
Shapiro~\cite{Gorenstein88}, Saha~\cite{Saha00}) and a correlation
between more compact mass distributions and longer time delays 
(e.g. Schechter~\cite{Schechter00}, Witt, Mao \& Keeton~\cite{Witt00}) 
were well known, it was unclear which properties of a halo model
had to be accurately computed in order to make robust predictions.
Kochanek~(\cite{Kochanek02b}) combined analytic results with 
comparisons to the numerical models by Kochanek~(\cite{Kochanek02a}) to
show that the surface density in the annulus between the images
used to measure the delay was the most important physical property
of the lens galaxy for determining the time delay.
The interior mass is implicit in the 
astrometry of the images and the lens galaxy, and the angular structure
is either unimportant or strongly constrained by the astrometry.
As a result, the Hubble constant expected for a simple lens is related to the
surface density by $H_0 = A(1-\kbar) + B(\eta-1)\kbar$, where 
$\kbar$ is the average surface density in the annulus between the
images (in units of the critical density), with a modest correction
$|B| \ltorder A/10$ due to the logarithmic slope $\eta$ of the surface 
density distribution within the annulus ($\kappa \propto R^{1-\eta}$).  
The coefficients $A$ and $B$ depend only on the image positions and
the measured time delay.  These simple semi-analytic scaling laws reproduce 
full numerical models to accuracies of better than 5\%.
  
We can now calculate the expected properties of gravitational lens time 
delays for CDM halo models.  In \S2 we outline our model for the halos,
which are based on the CDM lens models from Keeton~(\cite{Keeton01}). 
The models consist of a Hernquist~(\cite{Hernquist90}) model for the luminous 
early-type lens galaxy embedded in an NFW (Navarro, Frenk \& White~\cite{Navarro96}) 
halo normalized using the parameter estimates of Bullock et al.~(\cite{Bullock01}).
We considered both unmodified NFW halos and adiabatically compressed 
(Blumenthal et al.~\cite{Blumenthal86}) halos.  We summarize the mathematical
details of the model in \S2.   In \S3 we apply it to the four simple
time delay lenses PG1115+080, SBS1520+530, B1600+434 and HE2149--2745,
to show that the values of $\kbar$ and $\eta$ that determine the 
Hubble constant given the measured time delays are in turn determined by
a single parameter, the cold baryonic mass fraction, $f_b = \Omega_{b,cold}/\Omega_0$, 
of the luminous galaxy compared to the halo.  Since the baryon fraction is bounded by local
estimates from observed baryonic populations and the global baryon
fraction estimated either in clusters or from the CMB, we can set
firm bounds for the range of $H_0$ consistent with CDM halo models. 
As we discuss in \S4, this leads to a new element of the so-called
``dark matter crisis'' (e.g. Moore~\cite{Moore01}), because the CDM 
halo models combined with the measured time delays require lower
Hubble constants than are consistent with the Key Project estimates
based on the local distance scale.  An Appendix briefly discusses
the effects of tidal truncation on lens galaxy halos.
 
\def\bfx{{\bf x}}
\def\bfu{{\bf u}}
\def\grad{{\bf \nabla}}
\def\ka{\kappa_1}
\def\kb{\kappa_2}

\section{The CDM Lensing Model}

Our model for lensing by galaxies with standard CDM halos follows that of
Keeton~(\cite{Keeton01}).  These models are very similar to the lens
models used by Kochanek \& White~(\cite{Kochanek01b}) or those used
in semi-analytic models for the structure of galaxies 
(e.g. Mo, Mao \& White~\cite{Mo98}, Cole et al.~\cite{Cole00}, 
Gonzalez et al.~\cite{Gonzalez00} and references therein), but focus on early-type rather
than late-type galaxies.  We model the visible lens galaxy with a 
Hernquist~(\cite{Hernquist90}) profile,
\begin{equation}
    \rho_H(r) = { M_H \over 2 \pi } { r_H \over r (r+r_H)^3 },
    \label{eqn:hern1}
\end{equation}
with the scale length of $r_H=0.551 R_e$ that matches the Hernquist profile
to a de Vaucouleurs profile with effective radius $R_e$.  The enclosed
mass of the Hernquist model is
\begin{equation}
     M_H(<r) = M_H { r^2 \over (r+r_H)^2 }.
    \label{eqn:hern2}
\end{equation}
We model the
initial dark matter halo using an NFW (Navarro, Frenk \& White~\cite{Navarro96})
profile, 
\begin{equation}
    \rho_N(r) = { M_{vir} \over 4 \pi f(c) } { 1 \over r (r+r_s)^2 },
    \label{eqn:nfw1}
\end{equation}
normalized by the mass $M_{vir}$ inside the radius $r_{vir}$ with 
concentration $c=r_{vir}/r_s$ and $f(c)=\ln(1+c)-c/(1+c)$. The 
enclosed mass is 
\begin{equation}
   M_N(<r) = M_{vir} f(r/r_s)/f(c).
    \label{eqn:nfw2}
\end{equation}
We need the properties of these potentials in projection for the lensing
calculations.  If we define 
${\cal F}(x)=|x^2-1|^{-1/2} \hbox{tann}^{-1}|x^2-1|^{1/2}$, 
where the $\hbox{tann}^{-1} \rightarrow \tan^{-1}$ ($\hbox{tanh}^{-1}$) 
for $x >1$ ($x<1$), then the projected surface density of the Hernquist 
profile is
\begin{equation}
   \Sigma_H(x=R/r_H) = { M_H \over 2 \pi r_H^2 } 
       { (2+x^2){\cal F}(x)-3 \over (x^2-1)^2 } =  { M_H \over r_H^2 } \hat{\Sigma}_H(x)
    \label{eqn:hern3}
\end{equation}
and the mass inside cylindrical radius $R$ is
\begin{equation}
    M(<x=R/r_H) = M_H { x^2 \left( 1 - {\cal F}(x) \right) \over x^2 - 1 } = M_H \hat{M}_H(<x)
    \label{eqn:hern4}
\end{equation}
(Hernquist~\cite{Hernquist90}, Keeton~\cite{Keeton01}).
Similarly, the projected properties of the NFW model are
\begin{equation}
   \Sigma_N(x=R/r_s) = { M_{vir} \over 2 \pi r_s^2 f(c) } { 1 - {\cal F}(x) \over x^2 - 1}
        =  { M_{vir} \over r_s^2 }  \hat{\Sigma}_N(x)
    \label{eqn:nfw3}
\end{equation}
and 
\begin{equation}
   M_N(<x=R/r_s) = { M_{vir} \over f(c) } \left[ \ln { x\over 2} +  {\cal F}(x) \right]
      = M_{vir} \hat{M}_N(<x)
    \label{eqn:nfw4}
\end{equation}
for the projected surface density and mass respectively (Bartelmann~\cite{Bartelmann96}).
The lensing properties of the halo depend on the surface density measured in units
of the critical surface density, $\kappa=\Sigma/\Sigma_c$, where the critical surface
density, $\Sigma_c=c^2 D_{OS}/4\pi G D_{OL}D_{LS}$, is a simple function of the 
angular diameter distances between the observer, the lens and the source (e.g. Schneider,
Ehlers \& Falco~\cite{Schneider92}).

We need only two observational parameters to normalize the models.  The first is the
(intermediate axis) effective radius $R_e$ of the lens galaxy, which determines the density 
distribution of the visible baryons (stars). The second is the critical radius $R_c$
of the lens, determined either by detailed models or simply by the average distance
of the images from the lens center, which determines the mass scale. The halo properties
depend on the halo concentration, $c$, virial mass and radius, $M_{vir}$ and $r_{vir}$, 
and the mass fraction, $f_b$, represented by the luminous galaxy.  We adopt the halo
parameter estimates from Bullock et al.~(\cite{Bullock01}) and an $\Omega_0=0.3$
flat cosmological model.  The virial mass and radius are related by a definition,
\begin{equation}
     M_{vir} = { 4 \pi \over 3 } \Delta_{vir}(z) \rho_u(z) r_{vir}^3
             = 0.232 \left( { (1+z) r_{vir} \over 100 h^{-1}\hbox{kpc} } \right)^3
                 \left( \Omega_0 \Delta_{vir} \over 200 \right)
                 10^{12} h M_\odot
      \label{eqn:rvir}
\end{equation}
where $\rho_u(z)=3H_0^2\Omega_0(1+z)^3/8\pi G$ is the mean matter density at the lens 
redshift and the virial overdensity
is $\Delta_{vir}\simeq(18\pi^2+82x-39x^2)/\Omega(z)$ where $x=\Omega(z)-1$. 
Simulations find that the break radius and the virial radius are statistically
correlated through the concentration parameter $c=r_{vir}/r_s$.  The average
concentration depends on the halo mass $M_{vir}$ and the redshift $z$,  
\begin{equation}
     c = { 9 \over 1+z } \left( { M_{vir} \over 8.12 \times 10^{12} h M_\odot } \right)^{-0.14},
     \label{eqn:con}
\end{equation}
but individual halos have a log-normal dispersion about the average concentration
of approximately $\sigma_c=0.18$ (base 10).  Finally, a fraction $f_b$ of the
mass cools to form the visible galaxy, so the mass of the Hernquist profile
is $f_b M_{vir}$ and the remaining mass of the halo is $(1-f_b) M_{vir}$.  
  
We use the geometry of the lens to determine the overall mass of the halo.
The total projected mass inside the critical radius of a spherical
lens is $M(<R_c) = \pi R_c^2 \Sigma_c$ (e.g. Schneider et al.~\cite{Schneider92}), or 
\begin{equation}
     \pi R_c^2 \Sigma_c = 
       M_{vir} \left[ f_b \hat{M}_H(R_c/r_H) + (1-f_b) \hat{M}_N(R_c/r_s) \right]
      \label{eqn:rcrit}
\end{equation}
for a model without adiabatic compression.  We combine this equation with the relation
between the virial mass and radius (Eqn.~\ref{eqn:rvir}) to determine $M_{vir}$ and $r_{vir}$
given the cold baryon fraction $f_b$ and the concentration $c$.  The surface density,
in units of the critical density, is then
\begin{equation}
     \kappa(R) = { M_{vir} \over \Sigma_c} 
        \left[ { f_b \over r_H^2 } \hat{\Sigma}_H(R/r_H) + 
            { 1-f_b \over r_s^2 } \hat{\Sigma}_N(R/r_s) \right],
      \label{eqn:kappac}
\end{equation}
from which we can estimate both the average surface density $\kbar$ and the logarithmic
slope of the surface density $\eta$ (assuming $\kappa \propto R^{1-\eta}$ locally) 
in the annulus between the lensed images.

These models probably underestimate the central density of the dark matter because they neglect
the compression of the dark matter by the cooling of the baryons.  
We use the adiabatic compression model (Blumenthal et al.~\cite{Blumenthal86}) to compute 
the changes in the dark matter distribution.  For a spherical system with particles on
circular orbits, slow changes in the radial mass distribution preserve the angular
momentum of the particles.  For initial and final radii of $r_i$ and $r_f$ and
mass distributions $M_i(<r)$ and $M_f(<r)$, the angular momentum of a circular
orbit is conserved if
$ r_i M_i(<r_i) = r_f M_f(<r_f)$.  The initial mass distribution is simply that
of the NFW halo, $M_i(<r) = M_N(<r)$.  The final mass distribution is
\begin{equation}
      M_f(<r_f) = f_b M_H(<r_f) + (1-f_b) M_A(<r_f) = f_b M_H(<r_f) + (1-f_b) M_N(<r_i)  
      \label{eqn:adbat}
\end{equation}
where the final halo mass distribution $M_A(<r_f)=M_{vir}\hat{M}_A(<r_f)$ is the same as that of
the original halo at the initial radius $M_N(<r_i)$.  For the compression of
a NFW halo by Hernquist galaxy, the solution for $r_f(r_i)$ is analytic (a cubic 
equation, see Keeton~\cite{Keeton01}).  Jesseit et al.~(\cite{Jesseit02}) show
that simple Blumenthal et al.~(\cite{Blumenthal86}) model accurately estimates the
changes in the mass distribution. The compressed halo density profile must
then be numerically projected to determine the projected mass 
\begin{equation}
   M_A(<R) = M_{vir} \hat{M}_A(<R) = M_{vir} \int_0^{\pi/2} d\phi \cos \phi \hat{M}_A(<r=R\sec\phi) 
      \label{eqn:proj1}
\end{equation}
and surface density 
\begin{equation}
   \Sigma_A(R) = {M_{vir} \over r_s^2 } \hat{\Sigma}_A(R) = { M_{vir} \over 2 \pi R } 
     \int_0^{\pi/2} d\phi { d \hat{M}_A \over d r} (<r=R\sec \phi),
      \label{eqn:proj2}
\end{equation}
and then we must solve Eqns.~(\ref{eqn:rcrit}) and (\ref{eqn:kappac}) for $M_{vir}$
and $r_{vir}$
with the NFW profile replaced by the adiabatically compressed profile.  For one lens,
B1600+434, we include an exponential disk as well as a bulge in the baryonic model,
fixing the mass ratio to that inferred from the constant $M/L$ mass models of
Kochanek~(\cite{Kochanek02a}).  Once we have determined the surface density, we
can easily calculate the mean surface density $\kbar$ and the logarithmic slope
$\eta$ of the surface density in the annulus between the images for which time
delays have been measured.  

\begin{figure}
\centerline{\psfig{figure=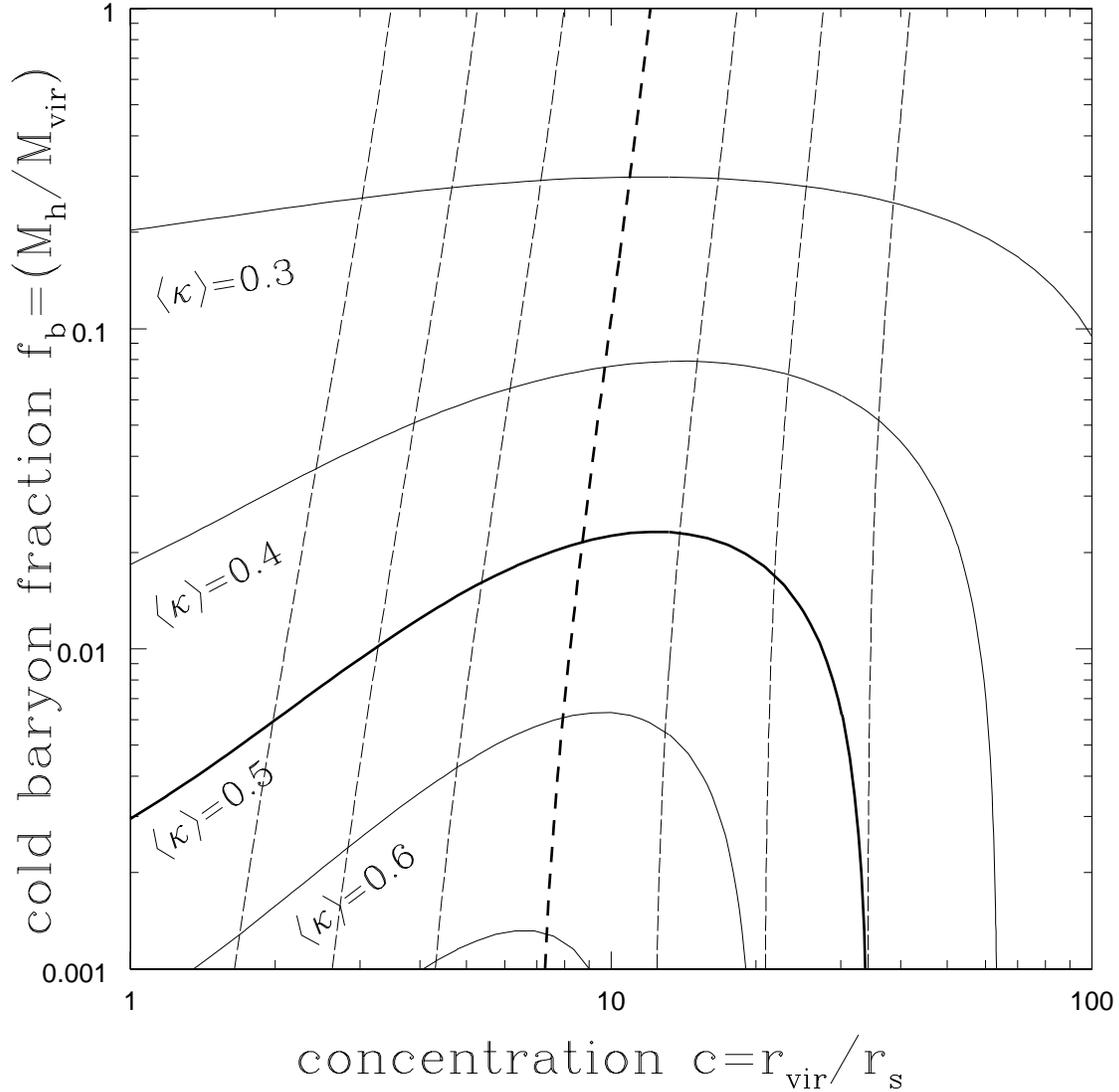,width=6.0in}}
\caption{ 
  The average surface density $\langle\kappa\rangle$ for adiabatically compressed models of 
  HE2149--2745 as a function of the halo concentration $c=r_{vir}/r_s$ and 
  the mass fraction in the cold baryons of the visible galaxy $f_b=M_H/M_{vir}$. 
  The solid lines are contours of $\langle\kappa\rangle$ spaced at intervals of 
  $\Delta\langle\kappa\rangle=0.1$. An isothermal model has $\langle\kappa\rangle=0.5$
  (the heavy solid contour).  Constant $M/L$ models lie on the top
  edge where $f_b=1$.  In the Bullock et 
  al.~(\protect{\cite{Bullock01}}) simulations of halo formation, halos have
  a limited range of concentrations for a given virial mass.  The heavy
  dashed contour shows the most likely halo concentration, and the light
  dashed contours show the 1, 2 and 3$\sigma$ ranges for the concentration
  distribution. 
  \label{fig:contourkappa}
  }
\end{figure}

\begin{figure}
\centerline{\psfig{figure=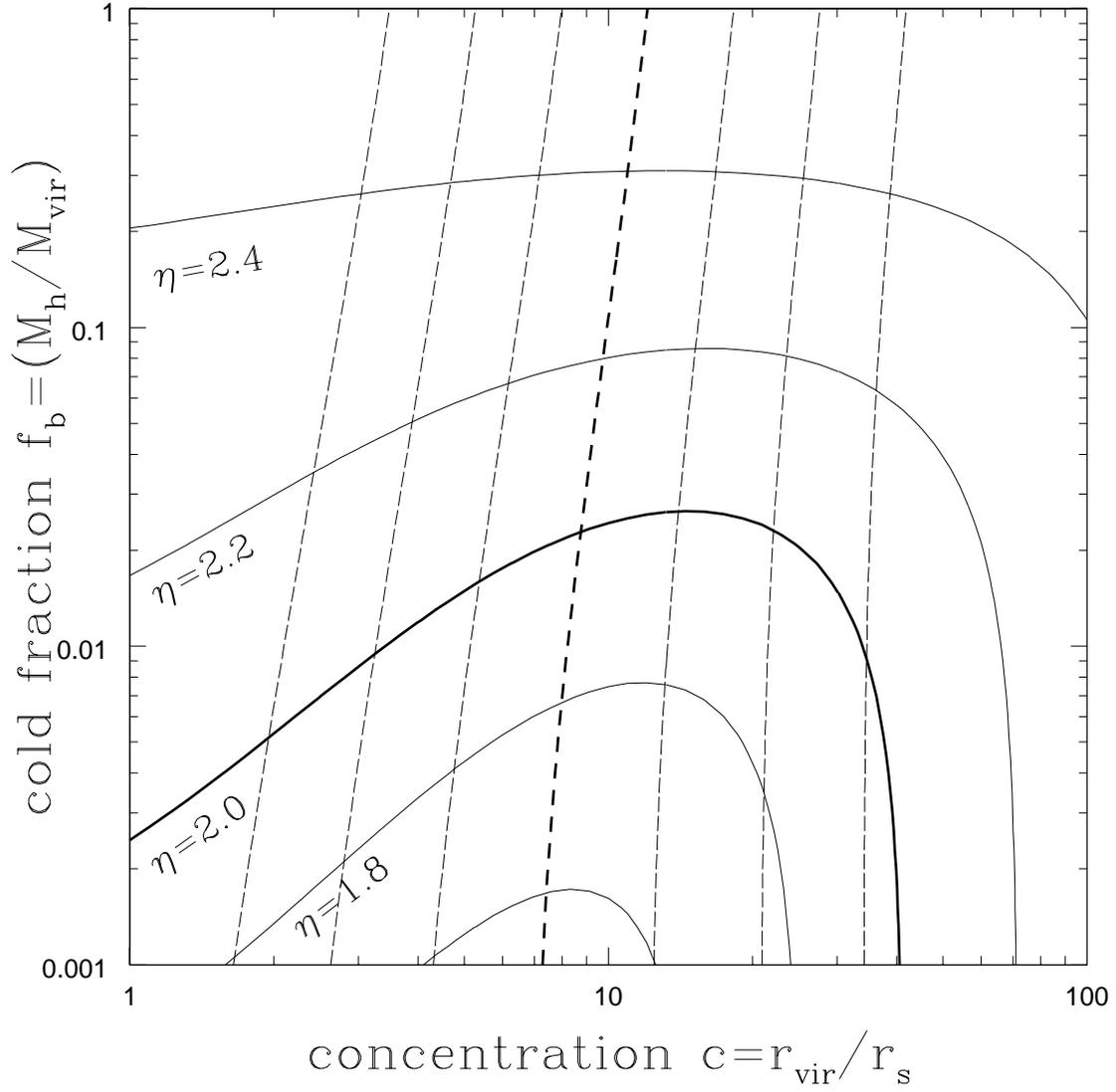,width=6.0in}}
\caption{
  The average logarithmic slope $\eta$ in the annulus between the images ($\kappa \propto R^{1-\eta}$) 
  for adiabatically compressed models of 
  HE2149--2745 as a function of the halo concentration $c=r_{vir}/r_s$ and 
  the mass fraction in cold baryons of the visible galaxy $f_b=M_H/M_{vir}$.
  The solid lines are contours of $\eta$ spaced at intervals of 
  $\Delta\eta=0.2$.  An isothermal model has $\eta=2$ (the heavy solid 
  contour).  The dashed contours show the likelihood of the concentration.
  The heavy dashed contour is the average concentration given the
  virial mass and the light dashed contours show the 1, 2 and
  3$\sigma$ ranges for the concentration (see Fig.~\protect{\ref{fig:contourkappa}}).
  \label{fig:contoureta}
  }
\end{figure}

\begin{figure}
\centerline{\psfig{figure=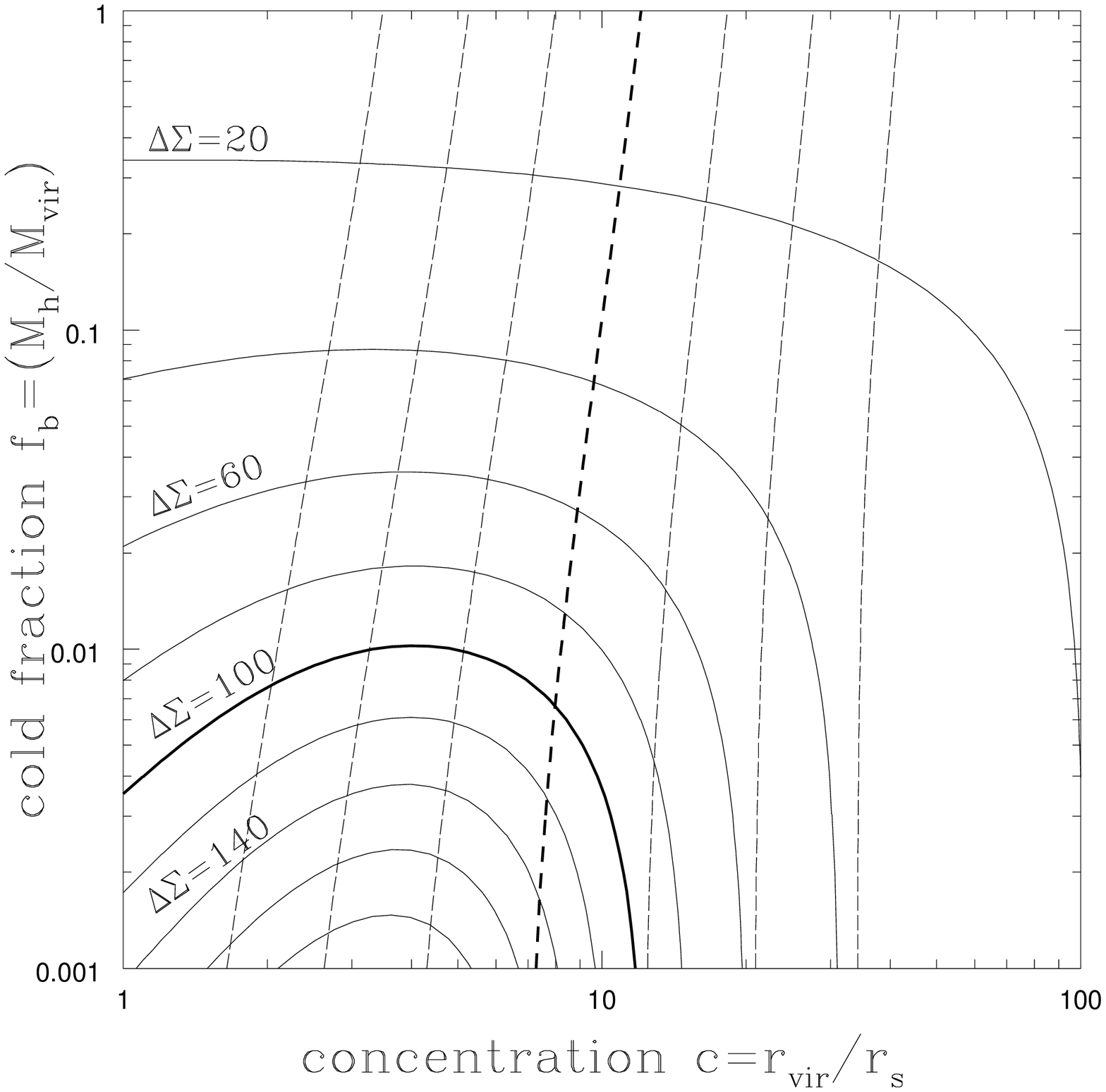,width=6.0in}}
\caption{
  The weak lensing surface density $\Delta\Sigma$ at $R_w=75h^{-1}$~kpc for
  HE2149--2745 as a function of the halo concentration $c=r_{vir}/r_s$ and
  the mass fraction in cold baryons of the visible galaxy $f_b=M_H/M_{vir}$.
  The solid lines are contours of $\Delta\Sigma$ spaced at intervals of
  $20hM_\odot$/pc$^2$ with a heavy contour for the surface density 
  expected from the SDSS weak lensing results, 
  $\Delta\Sigma_{SDSS}\simeq 100hM_\odot$/pc$^2$.
  An isothermal model would have 
  $\Delta\Sigma_{SDSS}\simeq 70hM_\odot$/pc$^2$.
  Based on the virial mass of each
  model, the dashed contours show the likelihood of the concentration.
  The heavy dashed contour is the average concentration given the
  virial mass and the light dashed contours show the 1, 2 and
  3$\sigma$ ranges for the concentration (see Fig.~\protect{\ref{fig:contourkappa}}).
  \label{fig:contoursigma}
  }
\end{figure}

In addition to estimates of the Hubble constant, we can also calculate the
weak lensing aperture mass, $\Delta\Sigma(R)=\langle \Sigma(<R)\rangle-\Sigma(R)$,
from the surface density profile.  Because weak lensing measurements are made 
on much larger scales ($R \gtorder 50h^{-1}$~kpc) than the critical radius of
the lens ($R \simeq 5h^{-1}$~kpc), they provide a nearly ideal additional 
constraint on the extent and mass of the dark matter halo.  Compact, low-mass,
constant $M/L$ models will produce smaller weak lensing signals than
extended, high-mass, dark matter dominated models.
While there are no weak lensing measurements for time delay
lenses, we can estimate the expected signal from the measurements by the 
Sloan Digital Sky Survey (SDSS, McKay et al.~\cite{McKay01}, Guzik \& Seljak~\cite{Guzik02}).
We consider only the innermost $R_w=75h^{-1}$~kpc bin of the SDSS measurements
in order to minimize contamination of the signal from sources other than the
galaxy halo.  In addition to any intrinsic uncertainties in the SDSS results,
the conversion of the results into a constraint on our models is complicated. 
We assume that we can estimate the central stellar velocity dispersion of the lens galaxy 
using an SIS lens model, which appears to be a fairly reliable assumption (see 
Kochanek et al.~\cite{Kochanek00}, Treu \& Koopmans~\cite{Treu02}).  We then use the 
observed properties of
early-type galaxies in the SDSS (Bernardi et al.~\cite{Bernardi02}) to convert the
velocity dispersion into estimates for the $r'$ and $i'$ magnitudes the lens 
galaxy would have locally.  This avoids any problems of luminosity evolution at 
the price of introducing the roughly 1~mag of scatter in the Faber-Jackson 
relation.  Combining these magnitudes with the $r'$ and $i'$ magnitude zero
points from Blanton et al.~(\cite{Blanton01}) and a fit to the $\Delta\Sigma(R)$ measurements 
in McKay et al.~(\cite{McKay01}), we estimate that the SDSS weak lensing results
correspond to a constraint on the aperture mass of
\begin{equation}
    \Delta\Sigma_{SDSS}(R_w=75h^{-1}\hbox{kpc}) 
   \simeq 80 \left( { \sigma \over 200\hbox{km/s} } \right)^{3.2}
               h M_\odot/\hbox{pc}^2
    \label{eqn:SDSS}
\end{equation} 
with a crudely estimated uncertainty of a factor of two.  That the weak lensing
signal is dominated by the galaxies has been checked by showing that the
dynamics of satellites around the target galaxies confirm the mass estimates
(McKay et al.~\cite{McKay02}).   Future analyses of weak lensing in the SDSS, particularly 
segregating the galaxies by velocity dispersion rather than luminosity, will clarify the 
scalings and the uncertainties. As a reality
check of this scaling, a singular isothermal sphere of velocity dispersion 
$\sigma$ predicts $\Delta\Sigma=62(\sigma/200\hbox{km/s})^2 hM_\odot$/pc$^2$,  
and suggests that the scaling of the SDSS results is somewhat high.
For now, the weak lensing results provide an interesting, but not a definitive,
constraint on the halo properties.

\section{Results}

We use the four simple, well-characterized time delay lenses PG1115+080
(Schechter et al.~\cite{Schechter97}, Barkana~\cite{Barkana97}, Impey et al.~\cite{Impey98}), 
SBS1520+530 (Burud~\cite{Burud02b},
Faure et al.~\cite{Faure02}), B1600+434 (Burud et al.~\cite{Burud00},
Koopmans et al.~\cite{Koopmans00}) and HE2149--2745 (Burud et al.~\cite{Burud02})
for our analysis, using the 
Hubble constant scalings derived in Kochanek~(\cite{Kochanek02b}).  Of the 
remaining 5 time delay lenses, B0218+357 lacks an accurate measurement
of the position of the lens galaxy (Lehar et al.~\cite{Lehar00}), B1608+656 has 
two interacting lens galaxies (Koopmans \& Fassnacht~\cite{Koopmans99}), Q0957+561 
and RXJ0911+0551 are too heavily perturbed by their parent clusters for our simple 
scaling with the surface density (see Kochanek~\cite{Kochanek02b}),
and the properties of the lens galaxy in PKS1830--211 are in
dispute (Winn et al.~\cite{Winn02}, Courbin et al.~\cite{Courbin02}). 
For each lens we model the baryonic distribution using the photometric
profiles scaled by their intermediate axis scale lengths (the geometric
mean of the major and minor axes).   Since the masses are determined by
the lens geometry, this is the only observational property of the lens
galaxy needed for our calculation.  The results are insensitive even to
large (50\%) changes in the scale length.  The only important uncertainties
in the measurements are the time delays and the B1600+434 lens galaxy
position.

Figures~\ref{fig:contourkappa} and \ref{fig:contoureta} show the expected values 
for $\kbar$ and $\eta$ in
HE2149--2745 as a function of the cold baryon fraction $f_b$
and the halo concentration $c=r_{vir}/r_s$.  For each value of $f_b$ and 
$c$, the virial mass (and radius) are determined by the observed critical
radius of the lens. As the cold baryon
fraction $f_b\rightarrow 1$, the surface density converges to that
of a constant $M/L$ model and the density exponent is relatively
steep ($\eta \simeq 2.5$).  The images lie somewhat outside the
effective radius of the lens ($R_e=0\farcs48$, $R_c=0\farcs85$,
$R_1=0\farcs34$ and $R_2=1\farcs35$) so the average slope of the
surface density is closer to the outer density exponent ($\eta=4$)
than the inner density exponent ($\eta=1$) of the Hernquist profile
(Eqn.~\ref{eqn:hern1}).
As we lower the cold baryon fraction $f_b$,
the expected surface density $\kbar$ rises and the slope $\eta$
flattens.  They reach the isothermal values, $\kbar=1/2$ and $\eta=2$,
for baryon fractions $f_b\simeq 0.025$.  Very extended halos
(small $c$) are too diffuse, and very compact halos (large $c$)
are too compact to contribute to $\kbar$. 

In simulations, halos with a fixed virial mass are found to have
a limited range of concentrations (see Bullock et al.~\cite{Bullock01} 
and Eqn.~\ref{eqn:con}), so only a portion of the $f_b$--$c$ 
plane can be occupied by real halos.  We find that the 
theoretically expected range of concentrations for the lenses
($c \simeq 10$) depends little on the cold baryon fraction, as
we can see in Figs.~\ref{fig:contourkappa} and \ref{fig:contoureta}
from the superposed contours for the expected range of concentrations
given by the Bullock et al.~(\cite{Bullock01}) results.  As a result, the
lensing properties depend almost exclusively on
the cold baryon fraction.  If we do not include the adiabatic
compression of the halo the results change slightly, but the
results are similar because the estimates with and without
compression must be the same
in the limits of both large and small baryon fractions. 

Figure~\ref{fig:contoursigma} shows the expected strength of the weak lensing
signal $\Delta\Sigma$ measured $R_w=75h^{-1}$~kpc from the lens.
The aperture mass/surface density which determines the measured weak lensing
shear steadily declines as we increase the cold baryon fraction and the total
halo mass decreases. This means that the
amplitudes of the shears measured by weak lensing provide a convenient 
observational constraint on the cold baryon fraction.  Implementing this
in practice is difficult, but for the scalings leading to Eqn.~(\ref{eqn:SDSS}) we
predict that $\Delta\Sigma_{SDSS}\simeq 100hM_\odot$/pc$^2$ for HE2149--2745
given the $214$~km/s velocity dispersion estimated from the image 
separations.  With our rough estimate of (logarithmic) uncertainties of
a factor of two, the discriminatory power of the weak lensing estimates
is limited but favors models with substantial amounts of dark matter.
Like the values for $\kbar$ and $\eta$, we can largely describe the 
weak lensing amplitude simply by the cold baryon fraction.

\begin{figure}
\centerline{\psfig{figure=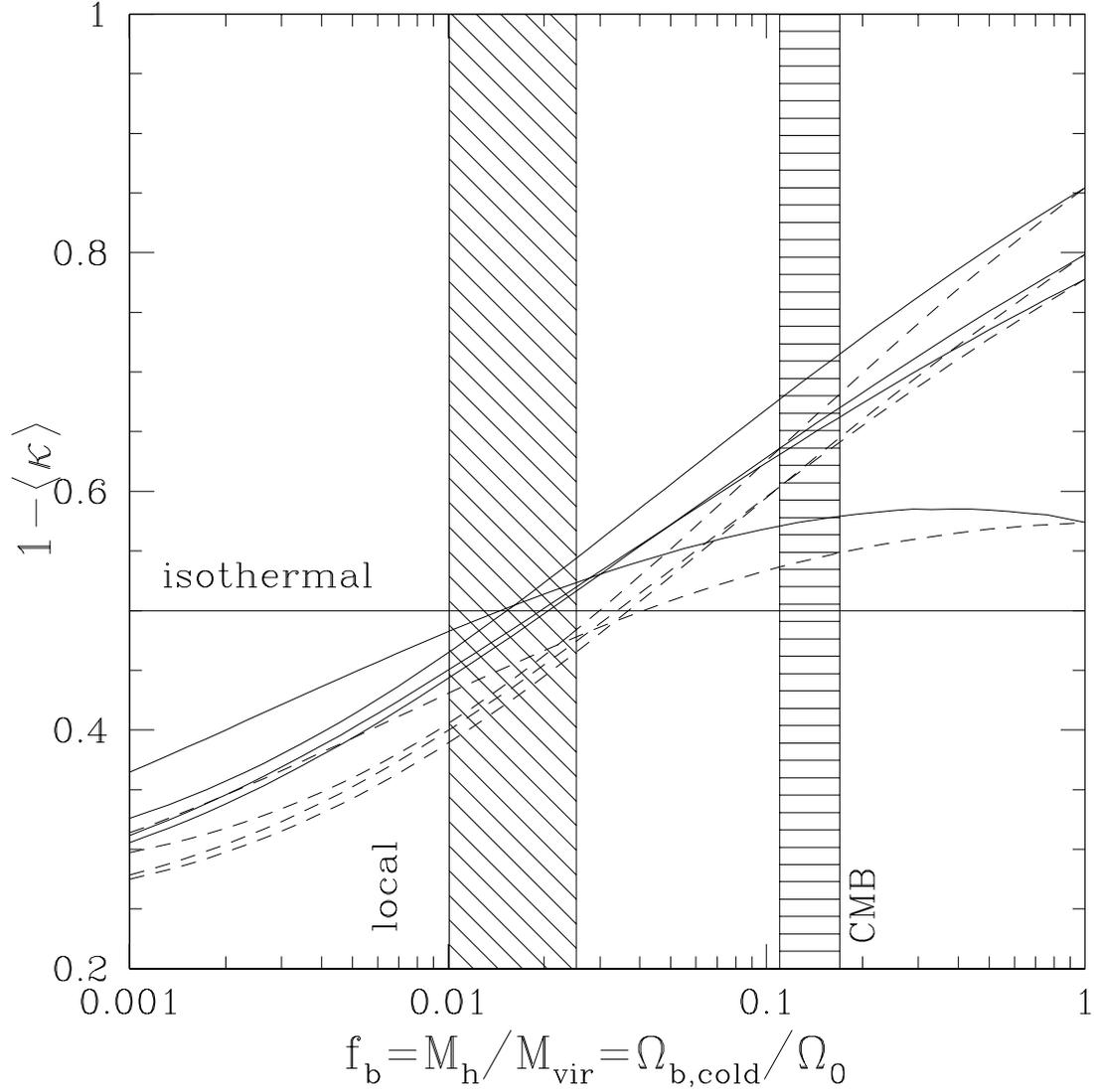,width=6.0in}}
\caption{
  The concentration-averaged estimates for $1-\langle\kappa\rangle$ as a function of $f_b$
  for PG1115+080, SBS1520+530, B1600+434 and HE2149--27145.  The solid (dashed) curves are 
  for models with (without) adiabatic compression.  
     A horizontal line shows the value expected for isothermal models.
  The diagonally cross-hatched region shows the lower bound on $f_b$ based on the local
  inventory of cold baryons by Fukugita et al.~(\protect\cite{Fukugita98}) for
  $H_0=60$~km/s~Mpc and $\Omega_0=0.3$.  The horizontally cross-hatched region
  shows the upper bound on $f_b$ set by the global ratio of $\Omega_b/\Omega_0$
  estimated from the CMB.
  \label{fig:plotkappa}
  }
\end{figure}

\begin{figure}
\centerline{\psfig{figure=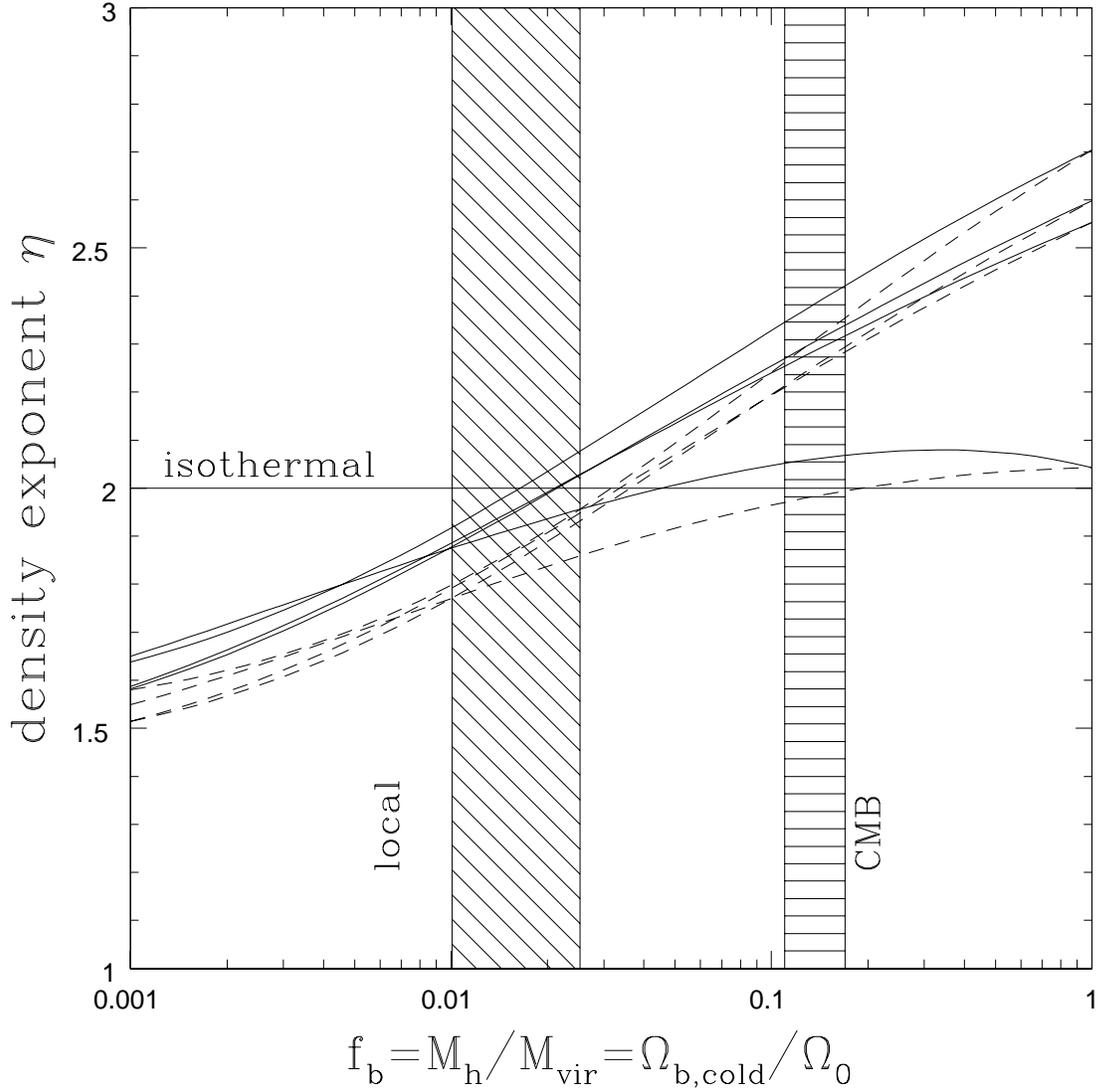,width=6.0in}}
\caption{
  The concentration-averaged estimates for the annular density exponent $\eta$ 
  ($\kappa \propto R^{1-\eta}$) as a function 
  of $f_b$ for PG1115+080, SBS1520+530, B1600+434 and HE2149--27145.  The solid (dashed) curves are 
  for models with (without) adiabatic compression.  
    A horizontal line shows the value expected for isothermal models.  
  The limits on $f_b$ are the same as in Fig.~{\protect\ref{fig:plotkappa}}.
  \label{fig:ploteta}
  }
\end{figure}

\begin{figure}
\centerline{\psfig{figure=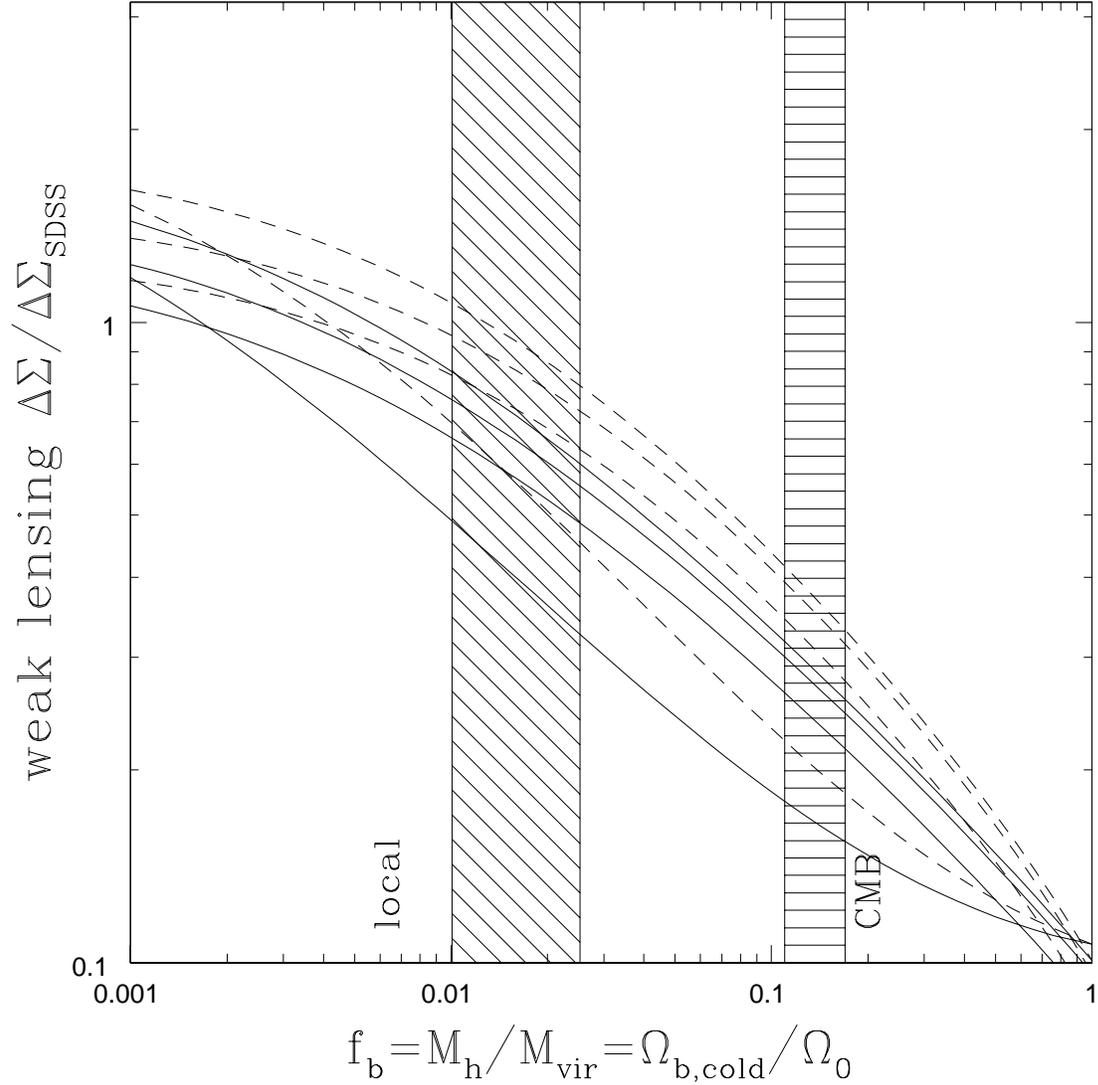,width=6.0in}}
\caption{
  The concentration-averaged estimates for the ratio $\Delta\Sigma/\Delta\Sigma_{SDSS}$ between the
  expected weak lensing signal $\Delta\Sigma$ and our best estimate of the SDSS measurements 
  $\Delta\Sigma_{SDSS}$ for galaxies with the velocity dispersion of the lens.  The normalization of 
  $\Delta\Sigma_{SDSS}$ is uncertain to a factor of 2 (0.3~dex). 
  The solid (dashed) curves are for models with (without) adiabatic compression. 
  The limits on $f_b$ are the same as in Fig.~{\protect\ref{fig:plotkappa}}.
  \label{fig:plotsigma}
  }
\end{figure}

\begin{figure}
\centerline{\psfig{figure=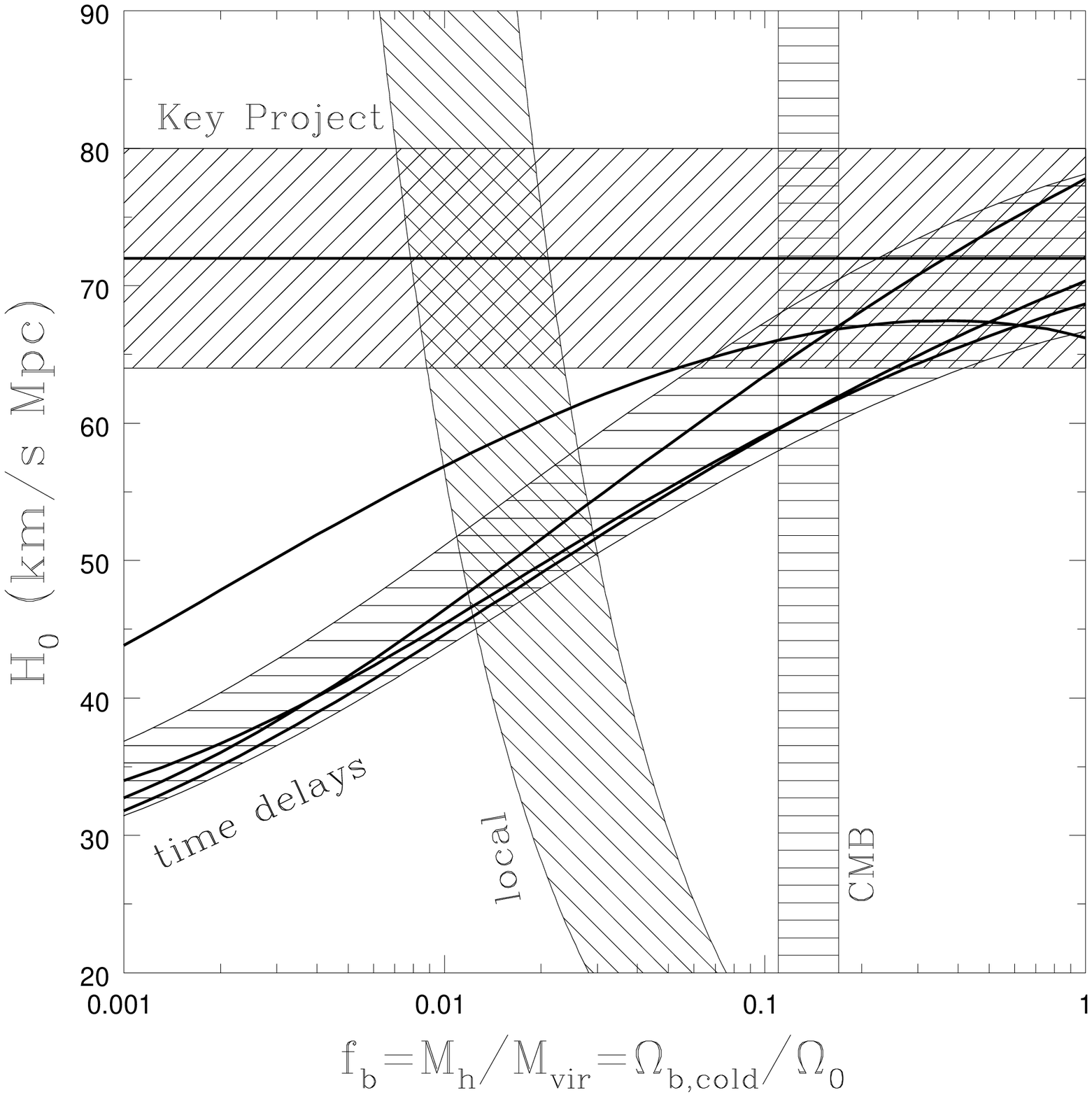,width=6.0in}}
\caption{
  The Hubble constant as a function of $f_b$ for the lenses PG1115+080, SBS1520+530,
  B1600+434 and HE2149--27145.  The heavy curves show the results for the individual lenses 
  including adiabatic compression.  The shaded envelope bracketing the curves is the 
  95\% confidence region for the combined lens sample. 
  The vertical bands show the lower bound on $f_b$ from
  local inventories, including its $H_0^{-1}$ scaling, and the upper bound from the
  CMB.  The horizontal band shows the estimate of $H_0=72\pm8$~km/s~Mpc by the HST
  Key Project (Freedman et al.~{\protect\cite{Freedman01}}).  Without adiabatic 
  compression, the estimates from the time delays predict values for $H_0$
  approximately 5~km/s~Mpc lower.
  \label{fig:ploth0}
  }
\end{figure}

The restricted range of permitted concentrations and the weak dependence
of the lens properties on the concentration in that range allows us to
simplify the results by considering only the concentration probability
averaged values for $\kbar$ and $\eta$ as a function of $f_b$. 
Figs.~\ref{fig:plotkappa} and \ref{fig:ploteta} show the 
expected values of $\kbar$ and $\eta$ as a function of the cold
baryon fraction for PG1115+080, SBS1520+530 B1600+434 and HE2149--2745.  
For a fixed cold baryon fraction $f_b$,
the results for the four lenses are similar, the differences
between adiabatically compressed and uncompressed models are small
and the scatter due to the permitted range of concentrations is
still smaller.  Over the range of baryon fractions from $f_b=10^{-3}$
to $f_b=1$, the expected surface density decreases from $\kbar\simeq 0.7$
to $\kbar\simeq 0.2$ and the exponent of the density profile steepens
from $\eta \simeq 1.6$ to $\eta\simeq 2.6$.  B1600+434 behaves differently
because it has a smaller ratio between the Einstein radius and the half-light
radius than the other three systems.

\def\omcold{\Omega_{b,\rm cold}}
The cold baryon fraction, $f_b=M_H/M_{vir}=\omcold/\Omega_M$, 
is set by the cosmological density $\omcold$ of baryons that
have cooled enough to be modeled by the visible galaxy compared
to the total density of baryons and dark matter $\Omega_M$.  
Local accountings for cold baryons (stars, remnants, cold
gas components) by Fukugita, Hogan \& Peebles~(\cite{Fukugita98}) 
estimated that $0.0024/h_{70} \ltorder \omcold \ltorder 0.0064/h_{70}$,
which sets a Hubble constant-dependent lower bound that
$f_b \gtorder 0.01$ for $\Omega_0=0.3$.  The global ratio
of $\Omega_b/\Omega_0$ sets an upper bound on the cold
baryon fraction.  If clusters of galaxies contain a fair 
sample of material (White et al.~\cite{White93}), their baryonic mass 
fraction determines 
$\Omega_b/\Omega_0 \simeq (0.113\pm0.005)(1+0.22h_{70}^{1/2})h_{70}^{-3/2}$
where the numerical estimate is from the recent study of
Allen, Schmidt \& Fabian~(\cite{Allen02}).
Current analyses of the cosmic microwave background (CMB)
anisotropies find $\Omega_b/\Omega_0\simeq 0.15\pm0.05$ 
(e.g. Netterfield et al.~\cite{Netterfield02}, 
Wang, Tegmark \& Zaldarriaga~\cite{Wang02}), consistent
with the cluster inventory results. In this context we
should also note that the CMB analyses generally favor lower
Hubble constants than the Key Project estimates.  
The local baryon inventory and the global ratio $\Omega_b/\Omega_0$
set upper and lower bounds on the cold baryon fraction $f_b$
appropriate for our models.  The cold baryon fraction could
be substantially higher than observed locally at the price
of introducing a cold but locally unobserved baryonic population.
We have superposed these local and global limits on $f_b$ on 
Figs.~\ref{fig:plotkappa} and \ref{fig:ploteta}.

For baryon fractions similar to local inventories ($f_b\sim 0.02$),
the surface density $\kbar$ and logarithmic slope $\eta$ are 
remarkably close to the isothermal values of $\kbar=1/2$ and
$\eta=2$.  For PG1115+080, SBS1520+530, B1600+434 and HE2149--2745
we find $\kbar=0.50\pm0.05$, $0.52\pm0.04$, $0.50\pm0.03$, and
$0.52\pm 0.04$ respectively, and $\eta = 1.98 \pm 0.09$,  
$1.95\pm0.09$, $1.91 \pm 0.05$, and $1.95 \pm 0.09$ respectively
if we constrain $f_b$ by the local values scaled to 
$H_0=60$~km/s~Mpc.  
This means that our standard isothermal lens model,
which represent our best independent estimate of the mass
distribution on these scales, is also the expectation from
standard CDM halo models.  If we allow larger cold baryonic
mass fractions, then the surface densities decrease and the
slopes steepen.  We find $\kbar = 0.30\pm0.04$, $0.34\pm0.03$,
$0.45\pm0.11$ and $0.35\pm0.03$, and $\eta=2.40\pm0.06$,
$2.32\pm0.07$, $2.02\pm0.27$ and $2.30\pm0.05$ for 
 PG1115+080, SBS1520+530, B1600+434 and HE2149--2745
respectively when we force $f_b=\Omega_b/\Omega_0=0.15\pm0.05$. 
These parameter values lie roughly midway between isothermal and
constant $M/L$ mass models.

The expected weak lensing signals, shown in Fig.~\ref{fig:plotsigma} as the
ratio $\Delta\Sigma/\Delta\Sigma_{SDSS}$ between our estimate from the models 
and our estimate for the SDSS measurements on a scale of $R_w=75h^{-1}$~kpc,
provides one independent test of the models.  The weak lensing signal
diminishes as $f_b$ increases and the total mass of the halo diminishes.
Models with $f_b$ similar to local inventories predict weak lensing 
signals roughly consistent with the SDSS measurements (about 70\% of
the expected amplitude, but we only believe the scaling to a factor of
two at present).  If all available baryons cool, the expected weak 
lensing signal is too weak to agree with the SDSS measurements (only
30\% of the expected amplitude).  Models with constant $M/L$ predict
weak lensing signals only 10\% of the expected amplitude and are
strongly ruled out.  With a factor of two uncertainty, the weak 
lensing constraint formally restricts the models to the range
$0.002 \ltorder f_b \ltorder 0.015$ (90\% confidence). Note that
when Keeton~(\cite{Keeton01}) used these lens models to fit the distribution
of image separations in lens surveys, he found it was difficult to match
the observed separation distributions for $f_b \ltorder 0.1$.

Finally, in Fig.~\ref{fig:ploth0} we use the estimates of $\kbar$ and $\eta$
for the CDM halo models to estimate the Hubble constant as a function of $f_b$ for 
PG1115+080, SBS1520+530, B1600+434 and HE1149--2745.\footnote{The scaling
solutions from Kochanek~(\protect\cite{Kochanek02b}) for these lenses 
are $H_0=A(1-\protect{\kbar})+B\protect{\kbar}(\eta-1)$ where the coefficients ($A$, $B$)
in units of km~s/Mpc for the lenses are ($92.3$, $4.6$), ($93.2$, $10.5$), 
($103.9$, $20.6$) and ($84.4$, $13.6$) for PG1115+080, SBS1520+530, B1600+434 
and HE1149--2745 respectively, and they produce fractional errors in $H_0$ 
due to uncertainties in the astrometry and the time delays of 9\%, 6\%, 9\%   
and 13\%, respectively, after broadening the errors in the time delays to
a minimum of 5\% to encompass systematic effects such as convergence
fluctuations from large scale structure (e.g. Seljak~\protect{\cite{Seljak94}},
Barkana~\protect{\cite{Barkana96}}). PG1115+080 is in a group which supplies
an additional external convergence of $\kappa_{ext}=0.2(1-\kbar)$.}
As we would expect
from the homogeneity of the lenses (see Kochanek~\cite{Kochanek02b}), these four lenses
predict similar values of $H_0$ as a function of $f_b$.  For low
$f_b$, B1600+434 gives somewhat higher values for $H_0$ than the 
other three lenses.
The joint estimate has small statistical
uncertainties (5\%) compared to the overall variable range.   
If the baryon density is restricted to agree with the local baryon
inventory, then $H_0=52\pm6$~km/s~Mpc (formally at 95\% confidence), 
and if it is restricted to agree with the global baryon inventory, 
then $H_0=65\pm6$~km/s~Mpc.  Restricted by the SDSS weak lensing
results rather than $f_b$, we find that $H_0=43\pm7$~km/s~Mpc,
but much of the lower range has baryon fractions below that found
in local inventories.  If we combine the weak lensing constraints
with the lower bounds on $f_b$ from the local inventories, we find
that $H_0=48\pm5$~km/s~Mpc.  
The limits for models without adiabatic compression are
approximately 5~km/s~Mpc lower.

Superposed on Fig.~\ref{fig:ploth0} is the local estimate of
$H_0=72\pm8$~km/s~Mpc by the HST Key Project 
(Freedman et al~\cite{Freedman01}).  The local estimate is
consistent with these four lenses only in the limit of a
constant $M/L$, as we had found previously 
(Kochanek~\cite{Kochanek02a}).  Formally, agreement with the Key
Project requires $f_b > 0.06$ (one-sided, 95\% confidence).
It is slightly inconsistent with models in which all baryons cool, 
and grossly inconsistent with models having realistic baryon
fractions or weak lensing signals consistent with the
SDSS measurements.

\section{Discussion}

Because gravitational lens time delays are determined by the Hubble
constant and the average surface density $\kbar$ of the lens galaxy
in the annulus between the images (Kochanek~\cite{Kochanek02b}), we
can make unambiguous estimates for the behavior of time delays in
standard CDM halo models.  In these models, the expected delay is
controlled by the mass fraction, $f_b=M_H/M_{vir}=\Omega_{b,cold}/\Omega_0$, 
in cold baryons making up the observed lens galaxy relative to the
overall halo.  As the cold
baryon fraction rises, so does the Hubble constant.  

When the cold baryon fraction is comparable to the local baryonic content
of galaxies ($f_b \simeq 0.02$, Fukugita et al.~\cite{Fukugita98}), the
model parameters closely match those for isothermal (flat rotation curve)
dark matter dominated lens models and the halos produce weak lensing
signals compatible with weak lensing measurements in the SDSS 
(McKay et al.~\cite{McKay01}). The mean surface density in the 
annulus is almost exactly $\kbar=1/2$ and the local slope of the
surface density is almost exactly $\eta=2$ ($\kappa \propto R^{1-\eta}$).  
Isothermal models are not only the best observational estimate for the
lensing potential on these scales, they are also the model predicted
by CDM assuming standard parameters and baryonic populations.
For baryon fractions with a lower limit
set by the local inventory and the upper limit set by the weak lensing
measurements, we find that $H_0=48\pm5$~km/s~Mpc based on the time
delays measured for PG1115+080, SBS1520+530, B1600+434 and HE2149--2745.
If all baryons were to cool and $f_b \simeq 0.15\pm0.05$, based on
constraints from either the CMB (e.g. Netterfield et al.~\cite{Netterfield02},
Wang et al.~\cite{Wang02}) or cluster baryon fractions (e.g.
White et al.~\cite{White93}, Allen et al.~\cite{Allen02}),
then the Hubble constant could be as high as $H_0=65\pm 6$~km/s~Mpc.
Such models require most of the cold baryons in the lens galaxies to
be in a locally invisible population and correspond to mass distributions
less consistent with direct estimates.
Both of these possibilities are lower than the local estimates of 
$H_0=72\pm8$~km/s~Mpc from the HST Key Project (Freedman et al.~\cite{Freedman01}),
which agrees with the time delays of these four lenses only for mass
distributions with constant $M/L$ ratios.  Thus, our detailed models
for the expected properties of time delays in standard CDM halos agree
with our simple models in Kochanek~(\cite{Kochanek02a}), and we are 
faced with a conflict between CDM halo models, gravitational lens time
delays and the local distance scale.

While there is some room for error in the lens results, the mutual
agreement of the four simple, well-characterized time delay lenses  
and the simple relation between time delays, surface densities and the
Hubble constant makes it difficult to point to a weakness (Kochanek~\cite{Kochanek02b}).
The most important observational steps are to improve the accuracies
of the existing delay measurements and to expand the number of systems
with delay measurements.  If the homogeneity of the results for simple
lens systems, as compared to more complicated systems in clusters
or with interacting galaxies, continues, the case for the existence
of a conflict will become overwhelming.
Improved characterizations of the lenses, either to constrain the
mass distribution in the time delay lenses directly or to allow us
to include the five other time delay lenses, are also important, 
but depend on obtaining deeper HST imaging of the systems.  Other
constraints on the mass distributions such as weak lensing or the
stellar dynamical measurements of the lens galaxies can also help
to break any degeneracies.  In particular, estimates of the weak
lensing signal as a function of the stellar velocity dispersion rather
than luminosity would be excellent constraints on the halo extent in
time delay lenses.

The systematic uncertainties in the mass distribution can be minimized by
measuring time delays in lenses where the baryons dominate the mass and
there there is little difference between a constant $M/L$ model and a
model with dark matter.  This means measuring the time delay in very 
low redshift lens galaxies where the ratio of the critical radius to the
effective radius, $R_c/R_e \propto D_{OL}$ is small and the mass near
the Einstein ring is increasingly dominated by the baryons.  For example, 
models of Q2237+0305 at $z_l=0.04$ suggest that less than 10\% of the mass 
inside the Einstein ring of the lens can be dark, instead of the roughly 
50\% for typical models of 
higher redshift lenses (Trott \& Webster~\cite{Trott02}).  Unfortunately,
Q2337+0305 has shown no variability on the very short time scale of its 
expected delay, making it a poor candidate for measuring time delays. 
There is a certain irony to proposing that local galaxies, which 
might be incorporated in local distance scale studies, are the ideal 
time delay lenses, but it may also lead to a system where the local
and the ``cosmological'' distance scales can be compared directly.

\noindent Acknowledgments.  
CSK thanks D. Rusin, P. Schechter, U. Seljak, J. Winn and S. Wyithe for 
discussions and comments.  CSK is supported by the Smithsonian Institution and 
NASA ATP grant NAG5-9265.  

\appendix

\section{Tidal Truncation}

Many lens galaxies, including many of the time delay lenses, are group and cluster
members (e.g. Keeton, Christlein \& Zabludoff~\cite{Keeton00}).  Particularly when 
PG1115+080 appeared to be the exception among time delay lenses rather than the
rule, tidal truncation of its halo by the surrounding group was frequently invoked
as a possible explanation for what appeared to be an anomalously low $H_0$ estimate
for isothermal lens models (e.g. Impey et al.~\cite{Impey98},
Koopmans \& Fassnacht~\cite{Koopmans99}).  The idea is that a 
tidally truncated halo is more centrally concentrated and has a lower $\kbar$,
so it produces a higher estimate of $H_0$ for a fixed time delay.  However, in order to 
significantly raise $H_0$, the halo must have a tidal truncation radius comparable to 
the critical radius of the lens and the surface density of the truncating halo
must be kept low.  While orbital history, phase, projection effects and differences in 
the density distribution make the physics of tidal truncation considerably more 
complicated then the simple model we discuss, simple scaling laws for
tidal truncation suggest that this standard picture is physically implausible.

The first problem with tidal truncation is that it is extraordinarily difficult
to truncate the mass distribution of a lens galaxy on scales comparable to the
critical radius.  For two SIS lenses, the tidal radius of the less massive
lens is $R_t = R_p (b_0/b_p)^{1/2}$ when the projected separation is equal
to the true separation in three dimensions.  The ratio roughly equals the velocity dispersion
ratio of the two halos, $(b_0/b_p)^{1/2} \simeq \sigma_0/\sigma_p$, so for
the typical massive lens galaxy in a low mass group or cluster the tidal radius
$R_t \gtorder R_p/2$.  Truncating the lens galaxy near the Einstein ring is
almost impossible because the inequality $R_t < b_0$ implies a galaxy
inside the critical radius of the cluster, $R_p < (b_p b_0)^{1/2} < b_p$,
where it cannot produce the image geometries seen for isolated lenses.
In fact, weak lensing studies of galaxies in the cores of rich clusters have shown that
even in these high density environments the halos of the galaxies seem to
be truncated only on scales $\gtorder 20h^{-1}$~kpc far larger than the
$\ltorder 5h^{-1}$~kpc scales on which tidal begins to alter the time
delays significantly (Natarajan et al.~\cite{Natarajan01}).  

The second problem is that the reduction in $\kbar$ from tidally truncating the
primary lens must be accompanied by an increase in $\kbar$ from the convergence
$\kappa_{ext}$ of the halo responsible for the truncation.  Existing lens
models partly incorporate this effect through the scaling of external shears
with surface density.  As a concrete example,
suppose we have a cruciform lens with images on the symmetry axes.  We consider
two models:  an SIS primary lens perturbed by an SIS cluster, and a point mass
primary lens perturbed by an SIS cluster. We model the cluster as an equal
external shear and convergence, $\gamma=\kappa=b_p/2R_p$, for a cluster with
critical radius $b_p$ at impact parameter $R_p$.  For a source directly behind
the lens, the two models can produce identical image positions.
The critical radius of the primary lens, $b_0$,
is the same in both models, but the cluster shear for a isothermal primary
is less than that for a point mass primary, with
$\gamma_{pnt}=\kappa_{pnt}=2\kappa_{SIS}(1-\kappa_{SIS})=2\gamma_{SIS}(1-\gamma_{SIS})$.  
As the lens galaxy becomes more compact, the external shear and convergence
from the cluster increase.  The time delay ratio
depends only on the strength of the cluster perturbation,
\begin{equation}
       {\Delta t_{pnt} \over \Delta t_{SIS} } =
          { 1-\kappa_{SIS} \over \kappa_{SIS} }
            \ln \left[ 1 \over 1 -2 \kappa_{SIS} \right]
           \rightarrow 2(1-\kappa_{SIS})+\cdots
\end{equation}
for small perturbations.  The factor of $2$ is the ratio of the $\kbar$ values
for a point mass and an isothermal primary lens, and the correction is due
to the higher convergence of the cluster when the primary lens is a point
mass.  The delay ratio is monotonically declining, reaching unity for
$\kappa_{SIS}=\gamma_{SIS}=0.36$ ($\gamma_{pnt}=0.45$).   Thus, as we
raise the cluster convergence, so as to make tidal truncation physically
plausible, the reduction in $H_0$ by the cluster convergence eventually
becomes more important than the increases in $H_0$ from making the
primary lens more centrally concentrated.  

While the rise in $\kappa_{ext}$ produced by the natural scaling of 
lens parameters with the increasing concentration of the lens galaxy
mimics the effects of tidal truncation, it does not necessarily lead 
to self-consistent results because the amplitudes of the shear
and convergence, $\gamma_{ext}$ and $\kappa_{ext}$, from the
perturbing halo are closely related to the tidal truncation radius.
For an isothermal galaxy truncated by an isothermal halo, the 
surface density of the perturber at the tidal radius is
$\kappa_{ext}=b_0/2 R_t \rightarrow 1/2$ as the tidal
radius approaches the critical radius, $R_t \rightarrow b_0$.   
Equivalently, there is a close relation between the tidal radius and the 
external shear from the perturbing halo, with 
$R_t = (b_0 b_p)^{1/2}/2\gamma_{ext}$ for two isothermal
halos, which means that the truncation radius cannot be
close to the Einstein ring unless the external shear is large.
Hence, the standard parameters of ``tidally truncated'' models of 
PG1115+080, where $\kappa_{ext}=\gamma_{ext}\simeq 0.15$ are 
inconsistent with simple models of the group parameters required
to tidally truncate the lens galaxy.  The standard model will
tend to overestimate $H_0$ for tidally truncated galaxies
by underestimating $\kappa_{ext}$.

\end{document}